\documentclass[%
 reprint,
superscriptaddress,
 amsmath,amssymb,
 aps,
prb,
]{revtex4-2}

\usepackage{graphicx, color}
\usepackage{dcolumn}
\usepackage{bm}
\usepackage{hyperref}
\hypersetup{
 citecolor=blue,
 colorlinks = true,
 urlcolor = blue
}
\usepackage[normalem]{ulem}
\usepackage{cancel}
\usepackage{physics}
\newcommand{\ee}{\mathrm{e}}
\allowdisplaybreaks[1]

\begin{document}

\title{Thermoelectric enhancement from an asymmetric spectral-conductivity cusp in spin-1 chiral fermions}

\author{Risako Kikuchi}
\affiliation{Department of Physics, Nagoya University, Nagoya 464-8602, Japan}

\author{Junya Endo}
\affiliation{Department of Physics, University of Tokyo, Bunkyo, Tokyo 113-0033, Japan}

\author{Ai Yamakage}
\affiliation{Department of Physics, Nagoya University, Nagoya 464-8602, Japan}

\date{\today}

\begin{abstract}
    A recent study showed that, in spin-1 chiral fermion systems composed of two linearly dispersing bands and one trivial band, impurity scattering produces an asymmetric cusp in the spectral conductivity $\sigma(\epsilon)$.
    We demonstrate that this asymmetric cusp markedly enhances the electronic thermoelectric response.
    Using linear-response theory within the self-consistent Born approximation, we find low-temperature enhancements in both the Seebeck coefficient and the electronic figure of merit.
    Increasing the curvature of the trivial band further strengthens this cusp-induced enhancement, even though the corresponding density of states becomes smoother.
    To clarify this mechanism, we introduce a minimal cusp model for $\sigma(\epsilon)$ and show that the enhancement is most pronounced when the cusp is sharp and strongly asymmetric, and when the spectral conductivity at the cusp energy is small.
\end{abstract}

\maketitle

\section{\label{sec:level1}Introduction}
Topological semimetals host symmetry-protected multiband crossings that give rise to emergent quasiparticles known as multifold fermions \cite{Ma2012, Beyond2016, Lv2021}.
In addition to the familiar spin-1/2 Dirac and Weyl fermions, which originate from two linearly dispersing bands, chiral crystals can realize threefold band degeneracies consisting of two linearly dispersing bands and an additional nearly flat trivial band.
The resulting quasiparticle is referred to as a spin-1 chiral fermion.
Three-dimensional realizations of spin-1 chiral fermions have been identified in chiral crystals with space group $P2_13$ (No.~198) \cite{Tang2017-kk, Chang2017, Pshenay-Severin_2018, Chang2018, Takane2019, Sanchez2019-by, Rao2019-ts, Schr2019, Li2019_RhSn, Mozaffari2020, Robredo_2024}.

Among the experimentally established hosts of spin-1 chiral fermions, transition-metal monosilicides crystallizing in space group $P2_13$ (No.~198) have attracted sustained attention.
A prototypical example is CoSi, which has long been studied not only as a topological semimetal but also as a room-temperature thermoelectric material~\cite{Asanabe1964, McNeill1964, Asanabe1965, Rowe1995, Kim2002, Lue2004, Ren2005, Li2005, Kuo2005, Sakai2007, Skoug2009, Sun2011, Sun2013, Pshenay2018_monosilicides, Pshenay2018_thermo, Burkov2018, Xia2019, Pshenay2019, Sk_2022, Nam2023, Ishibe2025, Zhong2025}.
In these systems, the coexistence of linearly dispersing bands and an additional trivial band produces a strongly energy-dependent density of states (DOS) near the Fermi level~\cite{Imai2001, Pan2007}, and sizable Seebeck coefficients have been reported, for example, $S \approx -80~\mu\mathrm{V/K}$ in CoSi~\cite{Ren2005, Sakai2007} and $S \approx -82~\mu\mathrm{V/K}$ in CoGe~\cite{Kanazawa2012} at 300~K.
Similar thermoelectric effects associated with coexisting linearly dispersing and trivial bands have also been discussed in RhSi~\cite{Nam2023}, the $\alpha$-T$_3$ lattice (a two-dimensional pseudospin-1 model)~\cite{Duan2023}, and Na$_2$AgSb, where an additional parabolic valence band passes through the linearly dispersing band at the Dirac point~\cite{Markov2019, Han2021}.
While these studies established the importance of this characteristic band structure for thermoelectricity, the present work addresses a different question: how impurity scattering affects the thermoelectric response in spin-1 chiral fermions.

Several quantitative studies have emphasized that reproducing the large near-room-temperature Seebeck coefficient in CoSi and related systems requires an energy-dependent relaxation time, particularly one associated with electron-phonon scattering~\cite{Pshenay2018_thermo, Xia2019, Nam2023, Ishibe2025}.
In contrast, we focus on the impurity-dominated low-temperature regime and clarify the scattering effects microscopically within the self-consistent Born approximation.

\begin{figure}
    \centering
    \includegraphics[width=\linewidth]{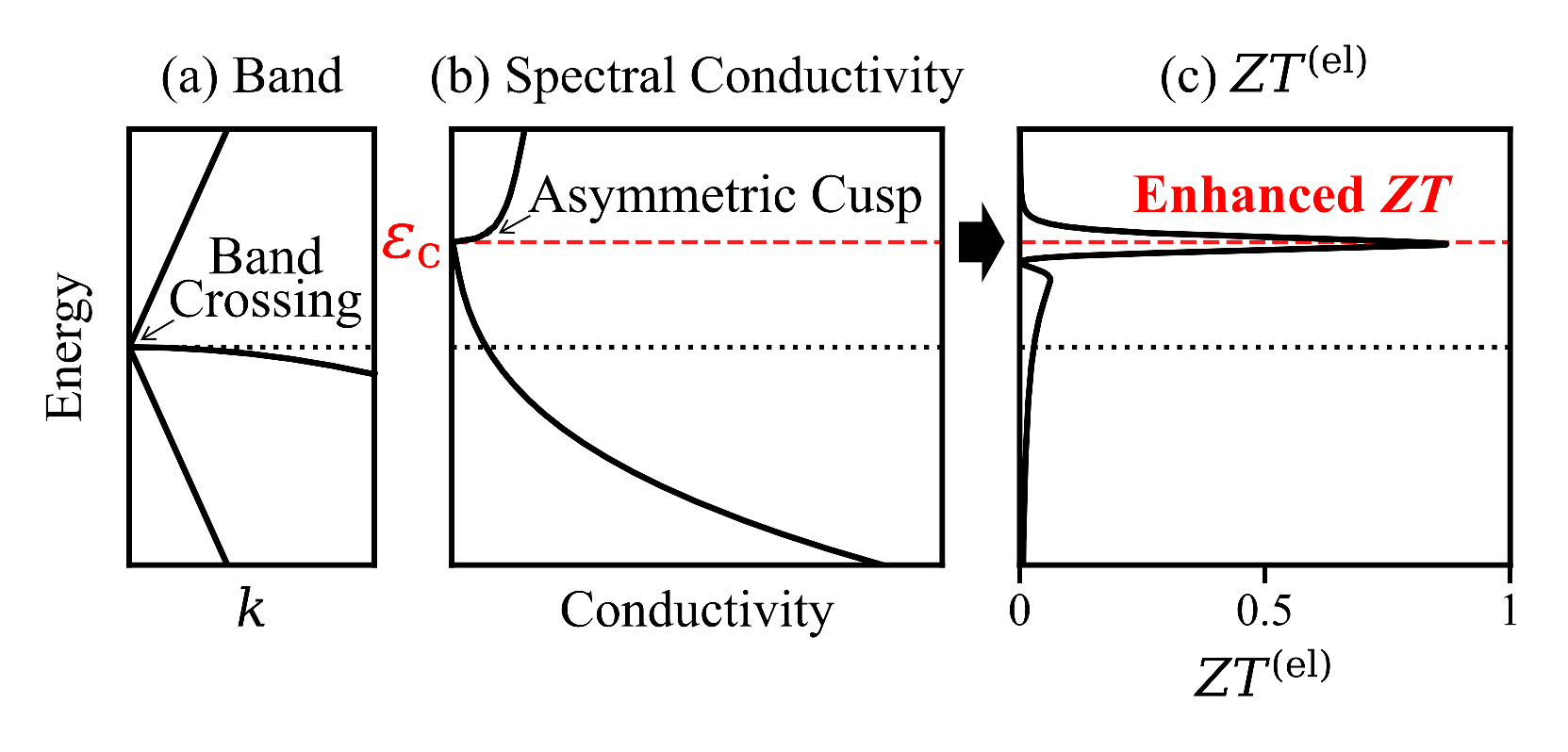}
    \caption{(Color online) Schematic overview of the physical picture and the main result of this work. (a) Band structure of an effective model for a threefold-degenerate spin-1 chiral fermion system. (b) Asymmetric cusp in the spectral conductivity at $\epsilon_{\mathrm{c}}$ (red dashed line)~\cite{Kikuchi2025}. (c) Electronic figure of merit $ZT^{(\mathrm{el})}$, showing an enhancement near $\epsilon_{\mathrm{c}}$ found in the present work.
    }
    \label{fig_intro}
\end{figure}

Quantum transport in band-crossing systems, such as Weyl systems, is well known to be highly sensitive to impurity scattering~\cite{Shon1998-ke, Noro2010-cm, Vigh2013, 0minato2014, Kobayashi2014, Nandkishore2014-vz, Ominato2015-um, Kikuchi2022, Kikuchi2023, Kikuchi2025, Leeb2023}.
This sensitivity manifests itself as a pronounced energy dependence with a cusp-like feature at $\epsilon_{\mathrm c}$ in the spectral conductivity (the $T=0$ electrical conductivity)~\cite{Shon1998-ke, Noro2010-cm, Vigh2013, 0minato2014, Ominato2015-um, Kikuchi2022, Kikuchi2023, Kikuchi2025, Leeb2023}.
For spin-1 chiral fermions, the resulting cusp is asymmetric in energy~\cite{Kikuchi2025}, as schematically shown in Figs.~\ref{fig_intro}(a) and \ref{fig_intro}(b).
Reference~\onlinecite{Kikuchi2025} characterized how the cusp energy $\epsilon_{\mathrm c}$ and the cusp conductivity $\sigma(\epsilon_{\mathrm c})$ vary with disorder strength and trivial-band curvature.
    Building on that result, we examine the thermoelectric consequences of such an asymmetric cusp in the low-temperature regime, where impurity scattering is expected to play a dominant role.
Indeed, the Mott formula~\cite{Mott1958} for the Seebeck coefficient $S$
\begin{equation}
    S \sim -\frac{\pi^{2} k_{\mathrm{B}}^{2} T}{3e}\,
    \left.\frac{\dd\ln \sigma(\epsilon)}{\dd\epsilon}\right|_{\epsilon=\mu},
    \label{eq:Mott}
\end{equation}
where $T$ is the temperature, $\sigma(\epsilon)$  is the spectral conductivity, $e>0$ is the elementary charge, and $k_{\mathrm{B}}$ is the Boltzmann constant, would imply an enhanced Seebeck coefficient in spin-1 chiral fermion systems. However, the cusp makes $\sigma(\epsilon)$ nondifferentiable at $\epsilon_{\mathrm c}$, and therefore the Mott formula cannot be applied straightforwardly.
This motivates a microscopic calculation of the thermoelectric response with impurity scattering.

In this study, we use linear-response theory~\cite{Luttinger1964}, with impurity scattering treated within the self-consistent Born approximation (SCBA), to investigate how the asymmetric cusp structure established for spin-1 chiral fermion systems in Ref.~\onlinecite{Kikuchi2025} affects their thermoelectric response.
Our aim is to clarify the mechanism by which cusp asymmetry in the spectral conductivity enhances the electronic thermoelectric response, using a single isotropic spin-1 node as a minimal model. Because we focus on impurity scattering, we evaluate only the electronic thermal conductivity and do not include the lattice contribution.
We find that spin-1 chiral fermion systems exhibit an enhanced low-temperature electronic thermoelectric response at energies where the asymmetric cusp forms, as schematically shown in Fig.~\ref{fig_intro}(c).
Moreover, we observe that the cusp-driven enhancement increases as the curvature of the trivial band becomes larger, even when the corresponding DOS variation becomes smoother.
To clarify the conditions for the enhancement of the electronic thermoelectric response, we further analyze a minimal effective model for the spectral conductivity with an asymmetric cusp structure.

The structure of this paper is as follows. Section~\ref{spin1} introduces the model for spin-1 chiral fermion systems and presents our analytical approach based on linear-response theory, along with results for their thermoelectric properties. Section~\ref{analysis} then examines a simplified spectral-conductivity model with cusp structures and clarifies their generic influence on thermoelectric effects. Building on these results, Section~\ref{discussion} compares the two analyses and identifies the mechanism underlying the nontrivial thermoelectric response caused by asymmetric conductivity cusps in spin-1 chiral fermion systems. Section~\ref{conclusion} summarizes our main results.

\section{Thermoelectric effect in a spin-1 chiral-fermion model}\label{spin1}
In this section, we study an effective model for spin-1 chiral fermion systems and derive low-temperature thermoelectric parameters within linear-response theory using the self-consistent Born approximation (SCBA). We obtain the density of states (DOS) and the electrical conductivity in SCBA, and then evaluate the Seebeck coefficient $S$, power factor $PF$, the figure of merit $ZT^{(\mathrm{el})}$, and the Lorenz ratio.

\subsection{Model}
We consider an isotropic, time-reversal-invariant three-dimensional spin-1 chiral fermion described by~\cite{Mandal2021-ic}
\begin{equation}
    \hat{\mathcal H}
    = \hbar v\,\hat{\boldsymbol S}\!\cdot\!\boldsymbol{k}
    + c\!\left[(\hat{\boldsymbol S}\!\cdot\!\boldsymbol{k})^{2}
        - k^{2}\hat S_{0}\right],
    \label{Hamiltonian}
\end{equation}
with $\boldsymbol{k}$ the momentum, $v$ the Fermi velocity, and $\hat S_{0}$ the $3\times3$ identity matrix.
The second term provides a quadratic-in-$k$ correction set by $c$.
The spin-1 generators are~\cite{Beyond2016}
\begin{equation}
    \hat S_x=\begin{pmatrix}0&i&0\\-i&0&0\\0&0&0\end{pmatrix},~
    \hat S_y=\begin{pmatrix}0&0&-i\\0&0&0\\i&0&0\end{pmatrix},~
    \hat S_z=\begin{pmatrix}0&0&0\\0&0&i\\0&-i&0\end{pmatrix}.
\end{equation}
The band energies are given by
\begin{align}
    \epsilon_{\mathrm c,\boldsymbol{k}} & =\hbar v k,  \\
    \epsilon_{\mathrm t,\boldsymbol{k}} & =-c k^2,     \\
    \epsilon_{\mathrm v,\boldsymbol{k}} & =-\hbar v k.
\end{align}

Disorder is modeled by finite-range (Gaussian) impurities,
\begin{equation}
    U(\boldsymbol{r})=\frac{\pm u_0}{(\sqrt{\pi}d_0)^{3}}\,\ee^{-r^{2}/d_0^{2}},
\end{equation}
with equal populations of $\pm u_0$ so that the Fermi level is unaffected by impurity concentration. The Fourier transform is
\begin{equation}
    u(\boldsymbol{k})=\!\int\! \ee^{-i\boldsymbol{k}\cdot\boldsymbol{r}}U(\boldsymbol{r}) {\dd}^{3}r
    =\pm u_0\,\ee^{-k^{2}/q_{0}^{2}},
    \
    q_0\equiv 2/d_0 .
\end{equation}
We quantify the scattering strength by
\begin{equation}
    W=\frac{q_{0}\,n_{\mathrm i}\,u_{0}^{2}}{\hbar^{2}v^{2}},
\end{equation}
where $n_{\mathrm i}$ is the impurity density, and introduce a dimensionless curvature parameter
\begin{equation}
    \tilde {c}=\frac{c q_{0}}{\hbar v}.
\end{equation}
We also define a reduced temperature
\begin{equation}
    \tilde{T}=\frac{k_{\mathrm B}T}{\hbar v q_0}.
\end{equation}

\subsection{\label{linear1}Sommerfeld--Bethe Relation}
The charge-current density $\boldsymbol{J}$ and heat-current density $\boldsymbol{J}_{\mathrm Q}$ respond linearly to an electric field $\boldsymbol{E}$ and a temperature gradient $\nabla T$ as
\begin{align}
    \boldsymbol{J}             & = L_{11}\,\boldsymbol{E} + L_{12}\!\left(-\frac{\nabla T}{T}\right), \\
    \boldsymbol{J}_{\mathrm Q} & = L_{21}\,\boldsymbol{E} + L_{22}\!\left(-\frac{\nabla T}{T}\right),
\end{align}
with transport coefficients $L_{ij}$ ($i,j=1,2$). Within the Sommerfeld–Bethe framework \cite{Ogata2019}, one may write
\begin{align}
    L_{11}                 & = \int \Bigl(-\frac{\partial {f(\epsilon)}}{\partial\epsilon}\Bigr)\,\sigma(\epsilon)\,\dd \epsilon, \label{eq:L11}                                      \\
    L_{12}=L_{21}          & = -\frac{1}{e}\int \Bigl(-\frac{\partial {{}f(\epsilon)}}{\partial\epsilon}\Bigr)(\epsilon-\mu)\,\sigma(\epsilon)\,\dd \epsilon, \label{eq:L12}          \\
    L_{22}^{(\mathrm{el})} & = \frac{1}{e^{2}}\int \Bigl(-\frac{\partial {{}f(\epsilon)}}{\partial\epsilon}\Bigr)\,(\epsilon-\mu)^{2}\,\sigma(\epsilon)\,\dd \epsilon, \label{eq:L22}
\end{align}
where $f(\epsilon)=[\ee^{(\epsilon-\mu)/(k_{\mathrm B}T)}+1]^{-1}$ is the Fermi–Dirac distribution, $\sigma(\epsilon)$ is the spectral conductivity, and ${L_{22}^{(\mathrm{el})}}$ denotes the electronic part of $L_{22}$.

From $L_{ij}$ we obtain the thermoelectric observables:
\begin{align}
    S                      & = \frac{1}{T} \frac{L_{12}}{L_{11}},                    \\
    \kappa^{(\mathrm{el})} & = \frac{L_{22}^{(\mathrm{el})} - L_{12}^2 / L_{11}}{T}, \\
    L                      & = \frac{\kappa^{(\mathrm{el})}}{L_{11} T},              \\
    PF                     & = L_{11} S^2,                                           \\
    ZT^{(\mathrm{el})}     & = \frac{L_{11} S^2}{\kappa^{(\mathrm{el})}} T.
\end{align}
Here, $L_{11}$ is the electrical conductivity, $S$ is the Seebeck coefficient,
$\kappa$ is the thermal conductivity, $L$ is the Lorenz number, $PF$ is the power factor, and $ZT$ is the figure of merit.
In this work, we evaluate the electronic figure of merit, $ZT^{(\mathrm{el})}$, using only the electronic thermal conductivity $\kappa^{(\mathrm{el})}$. The total figure of merit including the lattice (phonon) contribution, $ZT^{(\mathrm{tot})}$, would generally be smaller once the lattice thermal conductivity $\kappa^{(\mathrm{lat})}$ is taken into account.

\subsection{Linear Response Theory within SCBA}
For quenched disorder with a spatially random distribution of impurities, the impurity-averaged Green’s function reads
\begin{equation}
    \hat G(\boldsymbol{k},\epsilon+is0)=\bigl[\epsilon\,\hat S_{0}-\hat{\mathcal H}-\hat\Sigma(\boldsymbol{k},\epsilon+is0)\bigr]^{-1},
    \label{green function}
\end{equation}
where $s=\pm1$ labels retarded/advanced functions.
The self-energy satisfies the SCBA equation
\begin{equation}
    \hat\Sigma(\boldsymbol{k},\epsilon+is0)
    =\int\!\frac{n_{\mathrm i}\,\bigl|u(\boldsymbol{k}-\boldsymbol{k}')\bigr|^{2}}{(2\pi)^{3}}
    \,\hat G(\boldsymbol{k}',\epsilon+is0)\,\dd \boldsymbol{k}',
    \label{self energy}
\end{equation}
with impurity density $n_{\mathrm i}$ and impurity potential $u$. The density of states per unit volume is then
\begin{equation}
    D(\epsilon)=-\frac{1}{\pi}\,\Im\!\int\!\frac{\mathrm{Tr}\,\hat G(\boldsymbol{k},\epsilon+i0)}{(2\pi)^{3}}\,
    \dd \boldsymbol{k}.
    \label{dos}
\end{equation}

The spectral conductivity follows from the Kubo formula,
\begin{align}
    \sigma(\epsilon) & = -\frac{\hbar e^2 v^2}{4\pi}\sum _{s,s'=\pm1}ss'
    \notag                                                                               \\& \quad \times \int
    \text{Tr} \biggl[
        {\frac{\hat{v}_x}{v}}\hat{G}(\bm{k'},\epsilon+is0)
        \hat{J}_x(\bm{k'},\epsilon+is0,\epsilon+is'0)
    \notag                                                                               \\&\hspace{4em}\times
        \hat{G}(\bm{k'},\epsilon+is'0)
        \biggr]\frac{\dd \bm{k'}}{(2\pi)^3},\label{conductivity}
\end{align}
where $\hat v_{x}=(\hbar)^{-1}\partial \hat{\mathcal H}/\partial k_{x}$ and $\hat J_{x}$ is the dressed current vertex, determined self-consistently from the Bethe–Salpeter equation
\begin{align}
    \hat{J}_x(\bm{k},\epsilon,\epsilon') & =
    {\frac{\hat{v}_x}{v}} + \int\frac{n_{\text{i}}|u(\bm{k}-\bm{k'})|^2}{(2\pi)^3}\hat{G}(\bm{k'},\epsilon)\nonumber \\
                                         & \quad\times
    \hat{J}_x(\bm{k'},\epsilon,\epsilon')\hat{G}(\bm{k'},\epsilon')\,\dd \bm{k'}\label{Bethe}.
\end{align}
Technical details closely follow Ref.~\onlinecite{Kikuchi2025}.

The coupled self-consistency equations are solved numerically by iteration~\cite{Noro2010-cm}. To resolve the vicinity of the Dirac point, we employ a nonuniform radial mesh for $k$,
\begin{equation}
    dk_{j}=k_{\mathrm c}\,\frac{j}{\sum_{j=1}^{j_{\max}}j},\qquad
    k_{j}=\frac{dk_{j}}{2}+\sum_{j'=1}^{j-1}dk_{j'},
    \label{numerical}
\end{equation}
with $j=1,2,\dots,j_{\max}$, cutoff $k_{\mathrm c}$, and $j_{\max}=1000$, which concentrates points at small $k$ where singular structures are most pronounced.

\subsection{Results}\label{spin1-results}
Throughout this section, energies are expressed in units of $q_{0}\hbar v$. As an order-of-magnitude scale estimate, we take $q_0 \sim 0.01~\text{\AA}^{-1}$ \footnote{This value may be viewed as a Thomas–Fermi screening-wavevector estimate: taking a dielectric constant $\kappa \sim 10$ and an effective coupling $\alpha \sim 0.01$–$0.1$, one obtains $q_0 = \sqrt{(4\pi e^2/\kappa)\, D(\epsilon)} \sim 0.01~\text{\AA}^{-1}$ \cite{Kikuchi2023}. Because $D(\epsilon)$ is strongly energy dependent in the present system and our impurity model uses Gaussian finite-range disorder rather than a self-consistent screened-Coulomb treatment, this identification should be regarded only as an order-of-magnitude estimate.} and $v \sim 10^6~\text{m/s}$, we obtain $q_{0}\hbar v \simeq 0.066$ $\mathrm{eV}$.
The density of states (DOS), spectral conductivity, Seebeck coefficient, Lorenz number, and power factor are shown in units of $q_{0}^{2}/(\hbar v)$, $e^{2}q_{0}/\hbar$, $k_{\mathrm B}/e$, $(k_{\mathrm B}/e)^{2}$, and $k_{\mathrm B}^{2}q_{0}/\hbar$, respectively. With this normalization, the results do not explicitly depend on $q_{0}$.

\begin{figure*}[htbp]
    \centering
    \includegraphics[width=\linewidth]{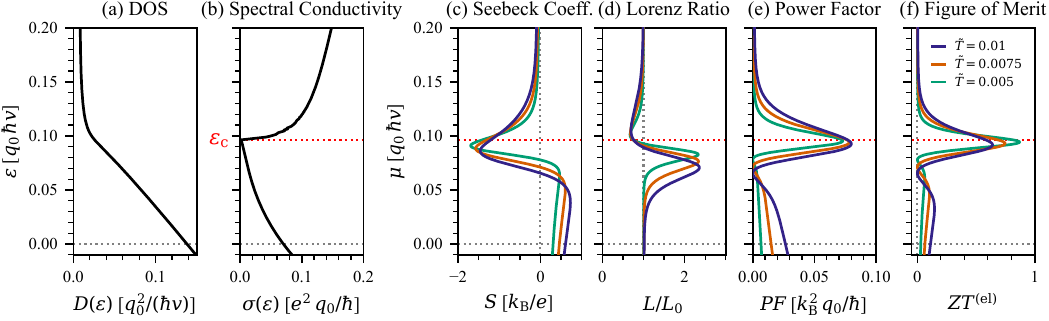}
    \caption{
        (Color online)
        Temperature dependences of thermoelectric response at fixed \(W=2\) and \(\tilde c=0.1\).
        (a) Density of states \(D(\epsilon)\) versus \(\epsilon\).
        (b) Spectral conductivity \(\sigma(\epsilon)\) versus \(\epsilon\).
        (c) Seebeck coefficient \(S\) versus chemical potential \(\mu\).
        (d) Lorenz ratio \(L/L_0\) versus \(\mu\).
        (e) Power factor \(PF\) versus \(\mu\).
        (f) Electronic figure of merit \(ZT^{(\mathrm{el})}\) versus \(\mu\).
        Panels (c)–(f) share a common color code for temperature: \(\tilde T=0.005\) (green), \(\tilde T=0.0075\) (orange), and \(\tilde T=0.010\) (indigo).
        In panels (b)–(f), the red dotted horizontal line marks \(\epsilon=\epsilon_{\mathrm c}\) in (b) and \(\mu=\epsilon_{\mathrm c}\) in (c)–(f).}
    \label{fig-T}
\end{figure*}
\begin{figure}[htbp]
    \centering
    \includegraphics[width=7cm]{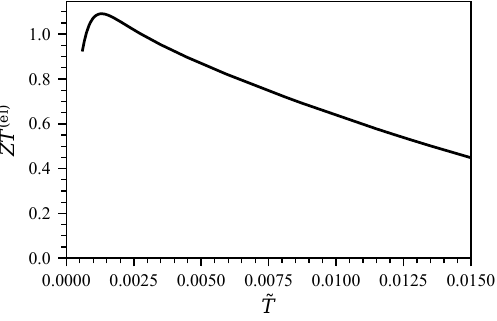}
    \caption{(Color online) Electronic figure of merit $ZT^{(\mathrm{el})}$ as a function of $\tilde{T}$ for the spin-1 chiral fermion model with $W=2$ and $\tilde c=0.1$, evaluated at the cusp chemical potential $\mu=\epsilon_{\mathrm c}$.}
    \label{fig-T2}
\end{figure}

Figure~\ref{fig-T} presents the results for $W=2$ and $\tilde{c}=0.1$.
The calculations are performed for $\tilde{T}=0.005, 0.0075,$ and $0.01$, which correspond to approximately $T\sim 3.82~\text{K}, 5.73~\text{K},$ and $7.64~\text{K}$ for the parameters assumed above.
The corresponding results for the electrical conductivity $\sigma$ and the thermal conductivity $\kappa$ are presented in Appendix~\ref{sec:spin1_conductivities}.
We focus on the peaks in $PF$ and $ZT^{(\mathrm{el})}$ at $\mu=\epsilon_{\mathrm c}\sim 0.096 q_0\hbar v$, at which an asymmetric cusp emerges in the spectral conductivity.
Around $\epsilon_{\mathrm c}$, the Seebeck coefficient [Fig.~\ref{fig-T}(c)] also exhibits a peak, and the Lorenz ratio $L/L_0$, with the Wiedemann–Franz value $L_0=(\pi^2/3)(k_{\mathrm B}/e)^2$, falls below unity [Fig.~\ref{fig-T}(d)],
resulting in an enhancement of both $PF$ and $ZT^{(\mathrm{el})}$.
As shown in Fig.~\ref{fig-T2}, $ZT^{(\mathrm{el})}$ exhibits a peak around $\tilde{T} \sim 0.00131$ and is suppressed at lower temperatures.
Such a temperature dependence is characteristic of this system; by contrast, in ordinary metals and semiconductors, the figure of merit $ZT^{(\mathrm{el})}$ decreases monotonically upon cooling.
This result demonstrates that the asymmetric conductivity cusp provides a mechanism for enhancing the electronic thermoelectric response at low temperatures in spin-1 chiral fermion systems. Whether this mechanism translates into a large total figure of merit $ZT$ depends on the lattice thermal conductivity, which lies beyond the scope of the present work.

\begin{figure*}[htbp]
    \centering
    \includegraphics[width=\linewidth]{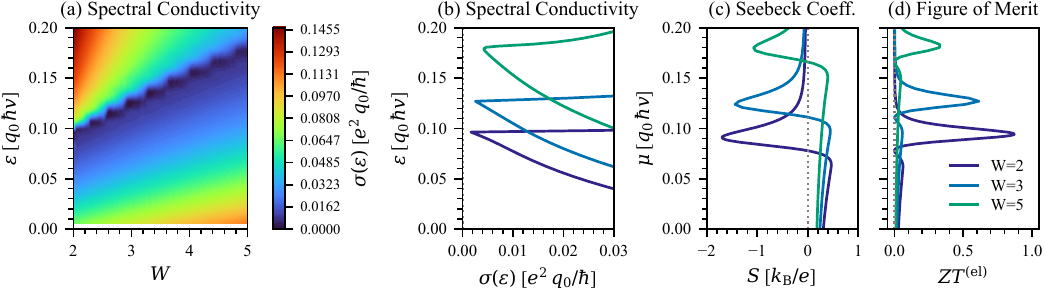}
    \caption{(Color online) Scattering-strength $W$ dependence of thermoelectric response at fixed $\tilde c=0.1$ and $\tilde T=0.005$.
        (a) Color map of the spectral conductivity $\sigma(\epsilon)$ in the $(W,\epsilon)$ plane.
        (b) $\sigma(\epsilon)$ versus $\epsilon$.
        (c) Seebeck coefficient $S$ versus chemical potential $\mu$.
        (d) Figure of merit $ZT^{(\mathrm{el})}$ versus $\mu$.
        Panels (b)–(d) share a common color code: $W=2$ (purple), $W=3$ (blue), $W=5$ (green).}
    \label{fig_W}
\end{figure*}

Figure~\ref{fig_W} shows the dependence on the scattering parameter $W$. The results are obtained with parameters fixed at $\tilde{c}=0.1$ and $\tilde T=0.005$. Increasing the scattering strength $W$ shifts the cusp energy $\epsilon_{\mathrm{c}}$ upward [Figs.~\ref{fig_W}(a) and \ref{fig_W}(b)]~\cite{Kikuchi2023}.
Correspondingly, the peaks in the Seebeck coefficient and figure of merit shift to higher energies and decrease in height [Figs.~\ref{fig_W}(c) and \ref{fig_W}(d)].
For $W=2$, the spectral conductivity at $\epsilon=\epsilon_{\mathrm c}$ is $\sigma(\epsilon_{\mathrm c})$ $\sim 0.001\, e^2 q_0/\hbar$. With $q_0 \sim 0.01~\textrm{\AA}^{-1}$, this value corresponds to $\sigma(\epsilon_{\mathrm c})$ $\sim 0.24~\textrm{S/cm}$.

\begin{figure*}[htbp]
    \centering
    \includegraphics[width=\linewidth]{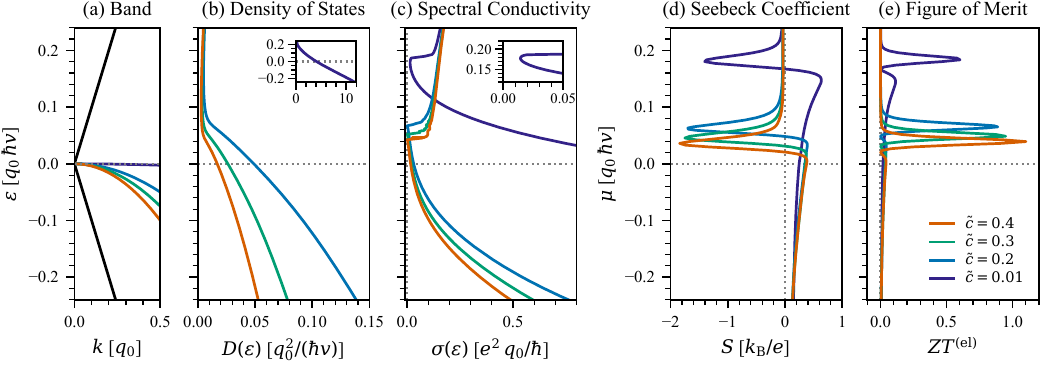}
    \caption{(Color online) Dependence on band curvature \(\tilde c\) at fixed \(W=2\) and \(\tilde T=0.005\).
        (a) Band dispersions \(\epsilon(k)\): two linearly dispersing bands} (black) and the trivial band for \(\tilde c=0.01,\,0.2,\,0.3,\,0.4\) (colored curves).
        (b) Density of states \(D(\epsilon)\) versus \(\epsilon\); inset: wide-range view of \(D(\epsilon)\) for \(\tilde c=0.01\).
        (c) Spectral conductivity \(\sigma(\epsilon)\) versus \(\epsilon\); inset: magnified low-\(\sigma\) region \(\bigl(0\le \sigma \le 0.05\,\mathrm{e}^{2} q_{0}/\hbar\bigr)\) and intermediate-energy window \(\bigl(0.12 \le \epsilon \le 0.22\, q_{0}\hbar v\bigr)\).
        (d) Seebeck coefficient \(S\) versus chemical potential \(\mu\).
        (e) Electronic figure of merit \(ZT^{(\mathrm{el})}\) versus \(\mu\).
        Panels (a)–(e) use a common color code for \(\tilde c\): \(0.01\) (indigo), \(0.2\) (blue), \(0.3\) (green), \(0.4\) (vermilion).
    \label{fig_c}
\end{figure*}

Figure~\ref{fig_c} shows the dependence on the curvature parameter $\tilde {c}$ of the trivial band. The results are obtained with the parameters fixed at $W=2$ and $\tilde{T} = 0.005$.
As the curvature increases, the density of states becomes smoother as a function of energy [Fig.~\ref{fig_c}(b)], while the asymmetry of the spectral-conductivity cusp becomes more pronounced [Fig.~\ref{fig_c}(c)].
Below $\epsilon_{\mathrm{c}}$, the spectral conductivity decreases more gradually, whereas above $\epsilon_{\mathrm{c}}$, it increases more steeply as a function of energy.
Correspondingly, the peaks in the Seebeck coefficient and figure of merit increase with curvature [Figs.~\ref{fig_c}(d) and \ref{fig_c}(e)].

In the next section, we present a more direct analysis of the relationship between the asymmetric cusp structure in the spectral conductivity and the enhancement of the figure of merit.

\section{Thermoelectric effects induced by a conductivity cusp}
\label{analysis}
In this section, we move beyond spin-1 chiral fermion systems and develop an effective spectral-conductivity model that captures generic cusp-like structures.
We aim to distill the minimal ingredients governing low-temperature thermoelectric responses and to clarify how cusp asymmetry impacts $S$, $PF$, $ZT^{(\mathrm{el})}$, and the Lorenz ratio.
We shift the energy origin to the cusp energy.

\subsection{Model and Formulation}
We focus on the low-energy regime in the vicinity of $\mu=0$ (i.e., $\epsilon\simeq 0$), and retain only the constant term and the leading $\epsilon$-dependent contribution in the spectral conductivity. We then introduce
\begin{equation}
    \sigma({\epsilon}) =
    \begin{cases}
        A\,\tilde{\epsilon}^{p} + \sigma_{\text{c}}            & (\tilde{\epsilon} \ge 0), \\[2pt]
        \gamma\,A\,(-\tilde{\epsilon})^{q} + \sigma_{\text{c}} & (\tilde{\epsilon} < 0),
    \end{cases}
    \label{eq:cusp_model}
\end{equation}
where $A$ is the coefficient of the $\tilde{\epsilon}^{p}$ term, $\sigma_{\text{c}}$ is the spectral conductivity at the cusp, $\gamma>0$ sets the amplitude ratio between the two sides, and $\tilde{\epsilon}=\epsilon/\epsilon_{0}$ is a dimensionless energy with an arbitrary scale $\epsilon_{0}$. In this work, we focus on the case in which the cusp represents a conductivity minimum and therefore take $A>0$ and $\sigma_{\text{c}}\ge 0$. The exponents $p,q>0$ control the cusp sharpness on the $\tilde{\epsilon}>0$ and $\tilde{\epsilon}<0$ sides, respectively; $\gamma=1$ and $p=q$ yield a symmetric cusp, while $p\neq q$ produces an asymmetric one (smaller exponents correspond to a sharper rise).
By normalizing with $A$ and defining $\tilde \sigma_{\text{c}}\equiv \sigma_{\text{c}}/A$, Eq.~\eqref{eq:cusp_model} becomes
\begin{equation}
    \frac{\sigma({\epsilon})}{A} =
    \begin{cases}
        \tilde{\epsilon}^{p} + \tilde \sigma_{\text{c}}            & (\tilde{\epsilon} \ge 0), \\[2pt]
        \gamma\,(-\tilde{\epsilon})^{q} + \tilde \sigma_{\text{c}} & (\tilde{\epsilon} < 0).
    \end{cases}
    \label{eq:cusp_model_normalized}
\end{equation}
Such low-energy parametrizations of energy-dependent transport functions are also used in thermoelectric analyses near the Anderson metal--insulator transition \cite{Croy2006}.

We define the integral function:
\begin{equation}
    g(n) = \int_0^\infty \frac{x^n \ee^x}{(\ee^x + 1)^2}\,\dd x
    = \int_{-\infty}^0 \frac{(-x)^n \ee^x}{(\ee^x + 1)^2}\,\dd x.
\end{equation}
This function can be expressed in terms of the Gamma function $\Gamma(n)$ and the Riemann zeta function $\zeta(n)$ as:
\begin{equation}
    g(n) =
    \begin{cases}
            1/2,
                                                         & \text{for } n = 0,     \\
            \ln 2,                                       & \text{for } n = 1,     \\
            \left(1 - 2^{1-n}\right)\Gamma(n+1)\zeta(n), & \text{for } n \ne 0,1.
        \end{cases}
\end{equation}
We then define the auxiliary quantities $F_m$, $m=0,1,2$, as
\begin{equation}
    F_m := \tilde{T}^{\,p} g(p+m) + (-1)^m \gamma\, \tilde{T}^{\,q} g(q+m),
\end{equation}
where $\tilde{T} := k_{\mathrm B} T/\epsilon_0$.

Using the Sommerfeld–Bethe relation at $\mu=0$, the transport coefficients are expressed as
\begin{align}
    \frac{1}{A}\,L_{11}(\mu=0)
     & = \int_{-\infty}^{\infty}
    \frac{\sigma(k_{\mathrm B}Tx)}{A}
    \frac{\ee ^{x}}{(\ee ^{x}+1)^{2}}\,\dd x
    \nonumber                                         \\&
    = F_0 + \tilde \sigma_{\text{c}}, \\
    \frac{e}{A}\,L_{12}(\mu=0)
     & =
    -k_{\mathrm B}T
    \int_{-\infty}^{\infty}
    \frac{\sigma(k_{\mathrm B}Tx)}{A}
    \frac{x\, \ee ^{x}}{(\ee ^{x}+1)^{2}}\,\dd x
    \nonumber                                         \\&
    = -F_1 k_{\mathrm B}T,                            \\
    \frac{e^{2}}{A}\,L_{22}(\mu=0)
     & =
    (k_{\mathrm B} T)^2
    \int_{-\infty}^{\infty}
    \frac{\sigma(k_{\mathrm B}Tx)}{A}
    \frac{x^{2}\ee ^{x}}{(\ee ^{x}+1)^{2}}\,\dd x
    \nonumber                                         \\&
    = \Bigl(F_2 + \frac{\pi^{2}}{3}\tilde \sigma_{\text{c}}\Bigr)
    (k_{\mathrm B}T)^{2}.
\end{align}
From these relations, the electronic thermoelectric response parameters are given as:
\begin{align}
    S                  & {= \frac{1}{T} \frac{L_{12}}{L_{11}}}
    =-\frac{k_\text{B}}{e} \frac{F_1}{F_0 + \tilde \sigma_{\text{c}}} , \label{eq:Cusp_S}                        \\
    L                  & {= \frac{L_{11}L_{22}^{(\mathrm{el})}-L_{12}^2}{T^2\,L_{11}^2}}
    = \left(\frac{k_\text{B}}{e}\right)^2 \frac{F_2 + ({\pi^2}/{3}) \tilde \sigma_{\text{c}}}{F_0 + \tilde \sigma_{\text{c}}} - S^2,
    \label{eq:Cusp_L}                                                                                                            \\
    PF                 & {= \frac{1}{T^2}\frac{L_{12}^2}{L_{11}}}
    = A\left( \frac{k_\text{B}}{e} \right)^2 \frac{F_{1}^2}{F_{0} + \tilde \sigma_{\text{c}}},\label{eq:Cusp_PF} \\
    ZT^{(\mathrm{el})} & {= \frac{L_{12}^2}{L_{11}L_{22}^{(\mathrm{el})}-L_{12}^2}}
    = \frac{F_1^2}{(F_0 + \tilde \sigma_{\text{c}})(F_2 + \frac{\pi^2}{3} \tilde \sigma_{\text{c}}) - F_1^2}\label{eq:Cusp_ZT}.
\end{align}

\subsection{Parameter Dependence of the Electronic Thermoelectric Response} \label{sec:cusp_results}

As follows from Eq.~(\ref{eq:Cusp_L}), when $\tilde{\sigma}_{\text c}$ is sufficiently large relative to $F_m$, the Wiedemann–Franz law holds approximately. As $\tilde{\sigma}_{\text c}$ is reduced, deviations from the Wiedemann--Franz law become significant. Moreover, Eqs.~(\ref{eq:Cusp_S}), (\ref{eq:Cusp_PF}), and (\ref{eq:Cusp_ZT}) show that smaller $\tilde{\sigma}_{\text c}$ enhances the Seebeck coefficient $S$, the power factor $PF$, and the figure of merit $ZT^{(\mathrm{el})}$, indicating that reducing $\tilde{\sigma}_{\text c}$ is favorable for the electronic thermoelectric response.

\begin{figure*}[htbp]
    \centering
    \includegraphics[width=\linewidth]{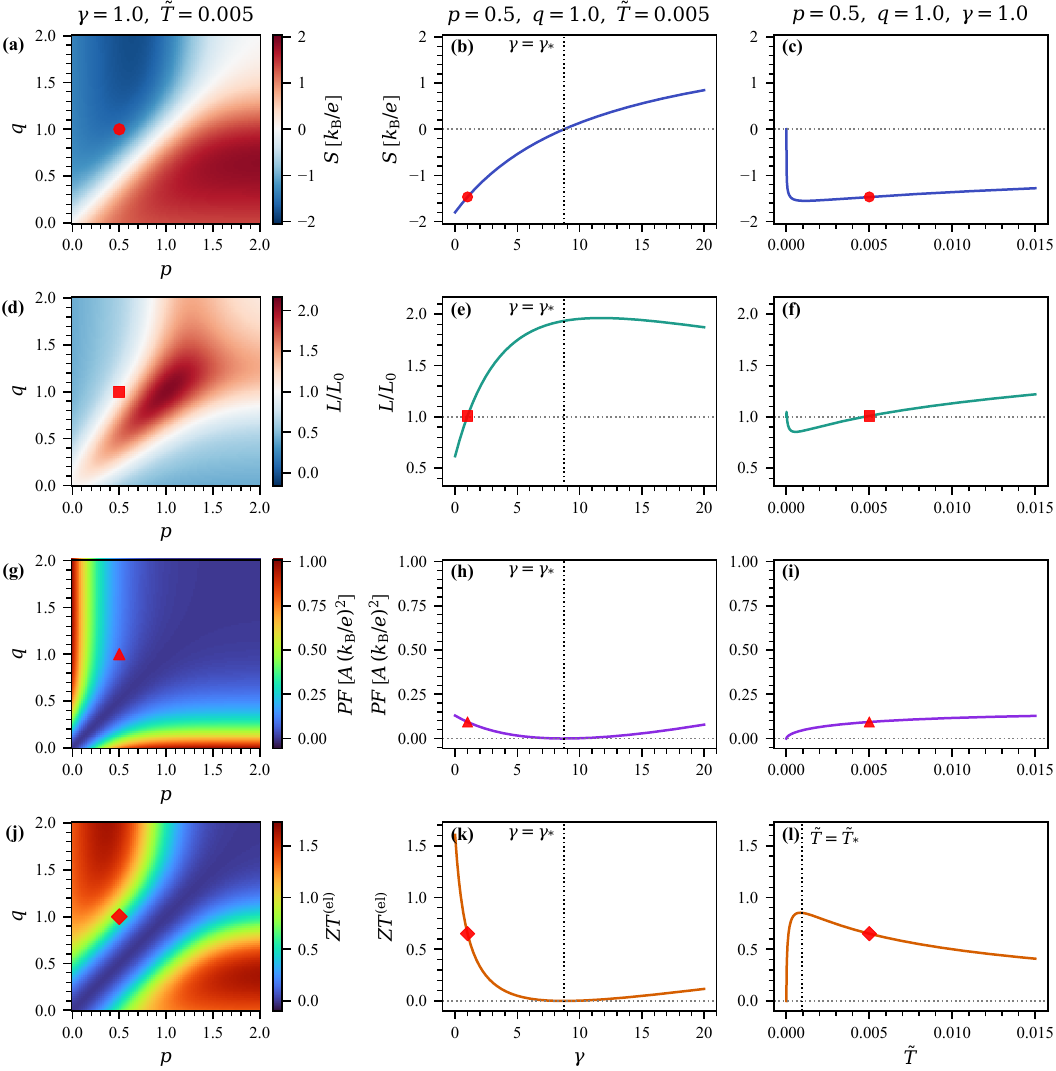}
    \caption{
        (Color online) Parameter dependence in the cusp model. Panels are arranged by quantity (rows) and control parameter (columns). We fix $\tilde{\sigma}_{\mathrm c}=0.002$ throughout.
        Rows (top to bottom): Seebeck coefficient $S$, Lorenz ratio $L/L_0$, power factor $PF$, and electronic figure of merit $ZT^{(\mathrm{el})}$.
        Left column [(a), (d), (g), and (j)]: color maps in the $(p,q)$ plane with $\gamma=1$ and $\tilde T=0.005$; the red dot marks the reference point $(p,q)=(0.5,1.0)$.
        Middle column [(b), (e), (h), and (k)]: dependence on $\gamma$ at fixed $p=0.5$, $q=1.0$, and $\tilde T=0.005$; the red marker indicates $\gamma=1$.
        Right column [(c), (f), (i), and (l)]: dependence on $\tilde T$ at fixed $p=0.5$, $q=1.0$, and $\gamma=1$; the red marker indicates $\tilde T=0.005$.
        The colormap for the Lorenz ratio is centered at $L/L_0 = 1$.
    }
    \label{fig_analysis}
\end{figure*}

Figure~\ref{fig_analysis} summarizes the parameter dependence of the cusp model. The left column presents color maps in the $(p,q)$ plane, and the middle and right columns show the dependence on $\gamma$ and on the reduced temperature $\tilde T$, respectively. Note that $PF$ is plotted in units of $A(k_{\mathrm B}/e)^2$, so the actual value of $A$ remains essential.
As seen in the left column of Fig.~\ref{fig_analysis}, $S$, $PF$, and $ZT^{(\mathrm{el})}$ vanish when $\gamma=1$ and $p=q$. Analytically, for $\gamma=1$ and $p=q$ one has $F_1=0$, so Eqs.~(\ref{eq:Cusp_S}), (\ref{eq:Cusp_PF}), and (\ref{eq:Cusp_ZT}) give $S=PF=ZT^{(\mathrm{el})}=0$. In other words, a symmetric cusp does not contribute to the thermoelectric response at the cusp energy.
Turning to the $\gamma$ dependence, the middle column of Fig.~\ref{fig_analysis} shows that $|S|$, $PF$, and $ZT^{(\mathrm{el})}$ are increased as $|\gamma-\gamma_*|$ is increased.
The value $\gamma_*$ that sets $S=PF=ZT^{(\mathrm{el})}=0$, equivalently, $F_1=0$, is
\begin{align}
    \gamma_*=\tilde T^{\,p-q}\frac{\,g(p+1)}{g(q+1)}.
\end{align}
This trend indicates that pronounced asymmetry,
characterized by a large $|\gamma - \gamma_*|$, enhances the electronic thermoelectric response.

When both exponents $p$ and $q$ are large, the cusp is smooth, and the left column of Fig.~\ref{fig_analysis} shows that $|S|$, $PF$, and $ZT^{(\mathrm{el})}$ are strongly suppressed and tend toward zero. In the low-temperature regime $\tilde T<1$, such large exponents make $\tilde{\sigma}_{\text c}$ dominant in the denominators of Eqs.~(\ref{eq:Cusp_S})--(\ref{eq:Cusp_ZT}); consequently, none of $|S|$, $PF$, or $ZT^{(\mathrm{el})}$ is enhanced, and $L$ approaches the Wiedemann--Franz value.
Hence, smooth cusps are unfavorable for low-temperature thermoelectric performance.
Note that too sharp a cusp, $p \ll q$ or $p \gg q$, is also unfavorable for low-temperature thermoelectric performance.
Appendix~\ref{sec:Detail_Cusp} further discusses how the $(p,q)$ region in which $|S|$, $PF$, and $ZT^{(\mathrm{el})}$ are strongly suppressed depends on the parameters $\tilde{\sigma}_{\mathrm c}$ and $\gamma$.

Finally, the right column of Fig.~\ref{fig_analysis} shows that, for $0<\tilde T<0.015$, the Lorenz ratio $L/L_0$ attains a minimum at small $\tilde T$, whereas $|S|$ and $ZT^{(\mathrm{el})}$ reach maxima. In Fig.~\ref{fig_analysis}(l), the maximum of \(ZT^{(\mathrm{el})}\) occurs at \(\tilde T \simeq 8\times10^{-4}\); using \(\tilde T = k_{\mathrm B}T/\epsilon_{0}\), this corresponds to \(T\simeq 0.9\,\mathrm{K}\) for \(\epsilon_{0}=0.1\,\mathrm{eV}\). The behavior of $ZT^{(\mathrm{el})}$ is qualitatively consistent with that in spin-1 chiral fermion systems (Fig.~\ref{fig-T2}). By contrast, $PF$ increases monotonically over the same interval.

Overall, these results suggest that, at low temperatures, $|S|$, $PF$, and $ZT^{(\mathrm{el})}$ tend to increase when the spectral conductivity at the cusp is small ($\tilde{\sigma}_{\mathrm c}$ is small), the cusp is strongly asymmetric ($|\gamma-\gamma_*|$ is large), and the cusp is sharp so that the cusp contribution is not masked by the constant background $\tilde\sigma_{\rm c}$.

\subsection{Approximate Evaluation of \texorpdfstring{$\tilde{T}_{*}$}{T*}}

We show analytically that the thermoelectric effect arising from the asymmetric cusp in the conductivity becomes pronounced at low temperatures, namely in the regime $\tilde T \ll 1$ ($k_{\mathrm B}T \ll \epsilon_0$), where the asymmetric cusp is not yet strongly smeared by thermal broadening.
To gain analytical insight into the temperature dependence of thermoelectric performance, we consider $\tilde T \ll 1$ with a small $\tilde \sigma_{\text c}$, and treat the exponents $p<q$ and the asymmetry $\gamma$ as fixed $O(1)$ parameters. Under these assumptions, we analyze $\dd(ZT^{(\mathrm{el})})/\dd\tilde T$ and, retaining only the leading contributions $\tilde T^{3p}$, $\tilde T^{2p+q}$, and $\tilde T^{2p}\tilde \sigma_{\text c}$, solve $\dd(ZT^{(\mathrm{el})})/\dd\tilde T=0$ approximately to obtain the temperature at which $ZT^{(\mathrm{el})}$ attains its maximum,
\begin{align}
    \tilde{T}_{*}(p, q, \gamma, \tilde \sigma_{\text c})
    \simeq \left(\frac{p\, G_{pq}}{(q - p)} \frac{\tilde \sigma_{\text c}}{\gamma}\right)^{1/q},
    \label{eq:Tmax}
\end{align}
where
\begin{align}
    G_{pq} = &
    g(p + 1)\,\bigl[g(p + 2) + \tfrac{\pi^2}{3}g(p)\bigr]\bigl[g(p + 1) g(p + 2) g(q)
    \nonumber  \\&
        + 2g(p) g(p + 2) g(q + 1)
    \notag     \\&
        + g(p) g(p + 1) g(q + 2)\bigr]^{-1}.
\end{align}
For $0<p<1$ and $0<q<2$, $G_{pq}$ is a dimensionless factor of order unity, with numerical values in the range $0.1\text{--}1$.
From this expression, $\tilde{T}_{*}$ becomes smaller as $\tilde \sigma_{\text c}$ is reduced, for fixed $p$, $q$, and $\gamma$. This suggests that reducing $\tilde \sigma_{\text c}$ both increases $ZT^{(\mathrm{el})}$ and shifts its maximum to lower temperatures.

\section{\label{discussion}Discussion}
\subsection{Physical origin of the cusp-driven thermoelectric enhancement in spin-1 chiral fermion systems}
\label{sec:discussion1}

The analysis in Sec.~\ref{analysis} shows that $|S|$ and $ZT^{(\mathrm{el})}$ increase at low temperatures when the spectral conductivity at the cusp is small and the cusp structure is both strongly asymmetric and sharp.
The behavior of the spin-1 chiral fermion system discussed in Sec.~\ref{spin1} can be understood from this perspective.
In this subsection, we interpret the numerical results presented in Sec.~\ref{spin1} in terms of the framework introduced in Sec.~\ref{analysis}.

First, we consider the effect of impurity scattering.
Increasing the disorder parameter $W$ makes the cusp smoother and raises the spectral conductivity at the cusp [Fig.~\ref{fig_W}(b)].
Within the framework of Sec.~\ref{analysis}, this corresponds to an increase in $\tilde{\sigma}_{\mathrm c}$ and a reduction of the cusp sharpness.
Both effects suppress the Seebeck coefficient $S$ and the electronic figure of merit $ZT^{(\mathrm{el})}$.
Thus, weaker disorder tends to yield larger $|S|$ and $ZT^{(\mathrm{el})}$.

Next, we examine the role of the trivial-band curvature.
Increasing the curvature parameter $\tilde c$ makes the cusp around $\epsilon_{\mathrm c}$ more asymmetric [Fig.~\ref{fig_c}(c)].
Within the framework of Sec.~\ref{analysis}, this corresponds to increasing $|\gamma-\gamma_*|$, which enhances $|S|$ and $ZT^{(\mathrm{el})}$.
Moreover, for small $\tilde c$, the system is more susceptible to impurity scattering.
As indicated by the $\tilde c=0.01$ curve in Fig.~\ref{fig_c}(c), impurity effects smooth the cusp structure, which further suppresses $|S|$ and $ZT^{(\mathrm{el})}$.
Thus, larger trivial-band curvature tends to increase $|S|$ and $ZT^{(\mathrm{el})}$.

Finally, we comment on the microscopic origin of the cusp structure.
As discussed in Ref.~\onlinecite{Kikuchi2025}, impurity-induced band mixing in this multiband system strongly suppresses the intraband contribution from the linearly dispersing bands to the spectral conductivity in the low-energy region.
Within the SCBA, the band crossing and the resulting band mixing play a crucial role in generating the sharp and strongly asymmetric cusp structure in the spectral conductivity around $\epsilon_{\mathrm c}$, which in turn contributes to the thermoelectric enhancement.

Taken together, these results suggest that weak impurity scattering and large trivial-band curvature favor a sharper and more strongly asymmetric cusp in the spectral conductivity, thereby increasing $|S|$ and $ZT^{(\mathrm{el})}$.
These results also highlight the importance of band crossing and the resulting band mixing in realizing such a cusp structure in the multiband system.

\subsection{Fitting Results}
In this section, we analyze the energy dependence of the spectral conductivity near the cusp for a spin-1 chiral fermion system with parameters $W=2$ and $\tilde c=0.1$.
We focus on $\sigma(\epsilon)$ around the cusp energy $\epsilon_\text{c}$ and fit the data using the following power-law functional form:
\begin{equation}
    \sigma(\epsilon) =
    \begin{cases}
        A(\tilde\epsilon - \tilde\epsilon_\text{c})^p + \sigma_\text{c},        & \tilde\epsilon \geq \tilde\epsilon_\text{c} \\
        A\gamma (\tilde\epsilon_\text{c} - \tilde\epsilon)^q + \sigma_\text{c}, & \tilde\epsilon < \tilde\epsilon_\text{c}
    \end{cases}
    \label{eq:fitting_func}
\end{equation}
\begin{table}
    \caption{Parameters in the cusp model for the spin-1 fermion system for $W=2$ and $\tilde c = 0.1$.}
    \begin{ruledtabular}
        \begin{tabular}{ccccccc}
            $\tilde\epsilon_{\rm c}$ & $\sigma_{\rm c} \, [e^2q_0/\hbar]$ & $\tilde\sigma_{\rm c}$ & $A \, [e^2q_0/\hbar]$ & $p$  & $q$  & $\gamma$
            \\ \hline
            0.096                    & 0.0017                                             & 0.0026                                 & 0.65                  & 0.47 & 0.94 & 0.53
        \end{tabular}
    \end{ruledtabular}
    \label{tab:param}
\end{table}

Here, $\epsilon_{\rm c}$ and $\sigma_{\rm c}$ are fixed from the numerical minimum, while $A$, $p$, $q$, and $\gamma$ are obtained by nonlinear least-squares fitting using scipy.optimize.curve\_fit.
Data within an energy range of $\epsilon_{\rm c} \pm 0.015 q_0 \hbar v$ are used for the fitting.
These parameters are summarized in Table~\ref{tab:param}.

\begin{figure}
    \centering
    \includegraphics[width=\linewidth]{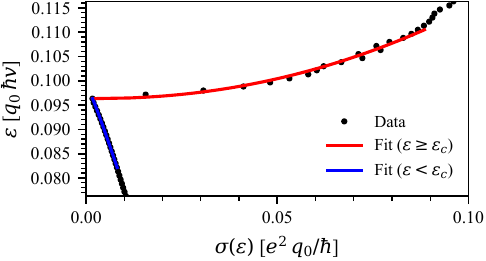}
    \caption{(Color online) Spectral-conductivity cusp fit by Eq.~(\ref{eq:fitting_func}) at fixed $W=2$ and $\tilde c=0.1$. Black circles: SCBA data for $\sigma(\epsilon)$. Red ($\epsilon\!\ge\!\epsilon_{\mathrm c}$) and blue ($\epsilon\!<\!\epsilon_{\mathrm c}$) lines: two-sided power-law fits,
    $\sigma(\epsilon)=A(\tilde\epsilon-\tilde\epsilon_{\mathrm c})^{p}+\sigma_{\mathrm c}$ for $\tilde\epsilon \ge \tilde\epsilon_{\mathrm c}$ and
    $\sigma(\epsilon)=A\,\gamma\,|\tilde\epsilon - \tilde\epsilon_{\mathrm c}|^{\,q}+\sigma_{\mathrm c}$} for $\tilde\epsilon < \tilde\epsilon_{\mathrm c}$,
    performed over $|\tilde\epsilon - \tilde\epsilon_{\mathrm c}|\le 0.015$ with $\tilde\epsilon_{\mathrm c}=0.096$ and $\sigma_{\mathrm c}=0.001658\,e^{2}q_0/\hbar$ fixed. Best-fit parameters are given in Table~\ref{tab:param} .
    \label{fig_fitting_result}
\end{figure}

Figure~\ref{fig_fitting_result} presents the fitting results alongside the data. The fitting curves (red for $\epsilon \ge \epsilon_\text{c}$ and blue for $\epsilon < \epsilon_\text{c}$) agree well with the numerical data.
Upon substituting these obtained values into Eq. (\ref{eq:Tmax}), we find $\tilde{T}_{*}\sim 0.00165$.
This result is of the same order as the temperature $\tilde{T} \sim 0.00131$ at which the spin-1 chiral fermion system reaches its peak $ZT^{(\mathrm{el})}$ under the same conditions ($\tilde c=0.1, W=2$), as discussed in Section~\ref{spin1-results}.

\section{\label{conclusion}Conclusion}
We show that asymmetric cusp structures at $\epsilon = \epsilon_{\mathrm c}$ in the spectral conductivity $\sigma(\epsilon)$ can enhance $|S|$ and yield appreciable $ZT^{(\mathrm{el})}$ at low temperatures.
The enhancement is governed by three ingredients: strong cusp asymmetry, sharp energy dependence on one side of the cusp, and a small cusp conductivity $\sigma_{\mathrm c} = \sigma(\epsilon_{\rm c})$.
Under these conditions, $\lvert S\rvert$ is enhanced and $ZT^{(\mathrm{el})}$ can become appreciable at low temperatures.

As shown in Ref.~\onlinecite{Kikuchi2025}, within the SCBA, such cusps arise from the multiband character of spin-1 chiral fermion systems.
To derive the asymmetric cusp structure in $\sigma(\epsilon)$ for spin-1 chiral fermions, it is essential to incorporate impurity-induced broadening of the trivial band near the band-crossing point; accordingly, we analyze the problem within the self-consistent Born approximation (SCBA).
Within this framework, we find that cusp asymmetry drives an increase in $\lvert S\rvert$ and $ZT^{(\mathrm{el})}$ at low temperatures.
The increase in $|S|$ and $ZT^{(\mathrm{el})}$ becomes more pronounced for weaker disorder and larger trivial-band curvature.

Altogether, these results identify a zero-field, disorder-driven mechanism for enhancing the low-temperature electronic thermoelectric response, rooted in the multiband character of spin-1 (multiply degenerate) chiral fermions and the energy asymmetry of their band structure.
They also suggest that asymmetric cusp structures in the spectral conductivity provide a useful perspective for understanding low-temperature electronic thermoelectric enhancement in multiband semimetals.

\begin{acknowledgments}
    This work was supported by JSPS KAKENHI (Grant Nos.\ JP25KJ1427, JP25K07224, and JP24H00853).
\end{acknowledgments}

\appendix

\section{Conductivities of a Spin-1 Chiral Fermion}
\label{sec:spin1_conductivities}
\begin{figure}[b]
    \centering
    \includegraphics[width=\linewidth]{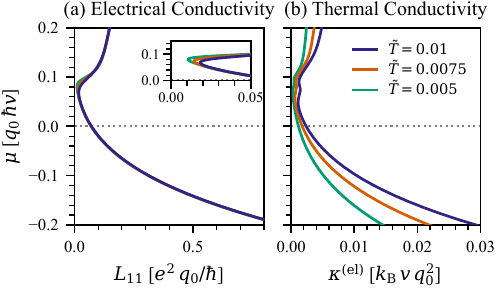}
    \caption{
        (Color online) Dependence on temperature at fixed \(W=2\) and \(\tilde c=0.1\).
        (a) Electrical conductivity \(L_{11}\) versus chemical potential \(\mu\); inset: magnified view of the low-\(L_{11}\) region \(\bigl(L_{11} \le 0.05\,\mathrm{e}^{2} q_{0}/\hbar\bigr)\) and low-\(\mu\) region \(\bigl(\mu \le 0.15\, q_{0}\hbar v\bigr)\).
        (b) Thermal conductivity \(\kappa^{(\mathrm{el})}\) versus \(\mu\); inset: magnified view of the low-\(\kappa^{(\mathrm{el})}\) region \(\bigl(\kappa^{\rm (el)} \le 0.005\,k_{\mathrm B} v q_0^2\bigr)\) and low-\(\mu\) region \(\bigl(-0.05 \le \mu \le 0.15\, q_{0}\hbar v\bigr)\).
        Panels (a) and (b) share a common color code: \(\tilde T=0.005\) (green), \(\tilde T=0.0075\) (orange), and \(\tilde T=0.010\) (indigo).
    }
    \label{fig_conductivities}
\end{figure}

Figure~\ref{fig_conductivities} shows the temperature dependence of the electrical and thermal conductivities in the spin-1 chiral fermion system. Both conductivities are strongly suppressed in the vicinity of $\epsilon_{\mathrm c} \sim 0.096 q_0 \hbar v$, leaving a clear remnant of the cusp. As temperature increases, the electrical conductivity becomes progressively smoother in energy, reflecting a thermal smearing of the cusp structure.

\section{Detailed Results of the Cusp Model}\label{sec:Detail_Cusp}
\begin{figure}
    \centering
    \includegraphics[width=\linewidth]{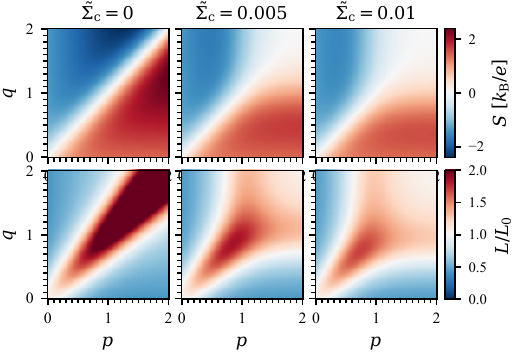}
    \caption{(Color online) Color maps of the Seebeck coefficient $S$ (top row) and the Lorenz ratio $L/L_0$ (bottom row) in the $(p,q)$ plane for several values of $\tilde{\sigma}_{\mathrm c}$.
        Throughout, $\gamma=1$ and $\tilde T=0.005$.
        For $L/L_0$, the colormap is centered at 1; the white region corresponds to $L=L_0$.
    }
    \label{fig_detail_analysis}
\end{figure}

In this appendix, we present further details of the parameter dependences in the cusp model introduced in Sec.~\ref{analysis}.
Figure~\ref{fig_detail_analysis} displays color maps of the Seebeck coefficient $S$ (top row) and the Lorenz ratio $L/L_0$ (bottom row) over the $(p,q)$ plane for several values of $\tilde{\sigma}_{\mathrm c}$.
As discussed analytically in Sec.~\ref{sec:cusp_results}, when $\tilde T^{\,p}\!\ll \tilde{\sigma}_{\mathrm c}$ and $\gamma\,\tilde T^{\,q}\!\ll \tilde{\sigma}_{\mathrm c}$ one has $S\to 0$ and $L$ approaches the Wiedemann–Franz value.
Consistent with this, Fig.~\ref{fig_detail_analysis} shows that, as $\tilde{\sigma}_{\mathrm c}$ increases $S$ decreases and $L$ approaches the Wiedemann–Franz value. Likewise, at fixed nonzero $\tilde{\sigma}_{\mathrm c}$, decreasing $\tilde T$ (within $\tilde T<1$) produces the same qualitative trend.

\begin{figure}
    \centering
    \includegraphics[width=\linewidth]{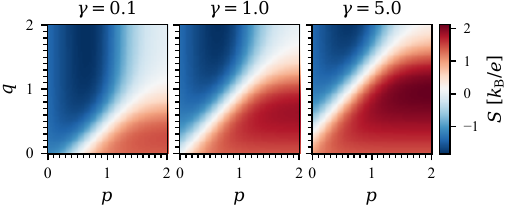}
    \caption{(Color online) Color maps of the Seebeck coefficient \(S\) in the \((p,q)\) plane for several values of \(\gamma\) (column titles). Throughout, \(\tilde T=0.005\) and \(\tilde{\sigma}_{\mathrm c}=0.002\). }
    \label{fig_detail_analysis2}
\end{figure}

Figure~\ref{fig_detail_analysis2} presents color maps of the Seebeck coefficient $S$ over the $(p,q)$ plane for several values of $\gamma$. As discussed analytically in Sec.~\ref{sec:cusp_results}, $S$ vanishes exactly when $\gamma=\tilde T^{\,p-q}\,g(p+1)/g(q+1)$.
Consistent with this relation, the $(p,q)$ maps show that the locus where $S=0$ aligns with $p=q$ for $\gamma=1$ and shifts systematically as $\gamma$ is varied.

\bibliography{ref}
\clearpage

\end{document}